\documentclass[preprint,12pt,3p]{elsarticle}

\usepackage{amssymb}
\usepackage{amsmath,mathtools}
\usepackage{comment}
\usepackage[svgnames]{xcolor}
\usepackage{soul} 
\usepackage{hyperref}
\usepackage{cuted}
\usepackage[normalem]{ulem} 
\biboptions{sort&compress}

\newcommand{\p}{\partial}
\newcommand{\erf}{{\rm erf}}
\newcommand{\ep}{\epsilon}

\catcode`\@=11
\def\gtsim{\mathrel{\vcenter{\m@th\offinterlineskip
\hbox{$\hfill>\hfill$}\kern.5ex\hbox{$\hfill\sim\hfill$}}}}
\catcode`\@=12

\catcode`\@=11
\def\ltsim{\mathrel{\vcenter{\m@th\offinterlineskip
\hbox{$\hfill<\hfill$}\kern.5ex\hbox{$\hfill\sim\hfill$}}}}
\catcode`\@=12

\hypersetup{colorlinks=true,linkcolor=green,filecolor=magenta,urlcolor=cyan,citecolor=cyan}

\journal{Combustion and Flame}

\begin{document}

\setstcolor{black}

\begin{frontmatter}

\title{Near-limit H$_2$-O$_2$-N$_2$ combustion in nonpremixed counterflow mixing layers}

\author[upm]{Jaime Carpio}
\author[ucsd]{Prabakaran Rajamanickam}
\author[ucsd]{Antonio L. S\'anchez}
\author[ucsd]{Forman A. Williams}

\address[upm]{Dept. Ingenier\'ia Energ\'etica, E.T.S.I. Industriales, Universidad Polit\'{e}cnica de Madrid, Madrid, 28006, Spain}
\address[ucsd]{Dept. Mechanical and Aerospace Engineering, University of California San Diego, La Jolla CA 92093-0411, USA}

\begin{abstract}
Numerical computations employing detailed chemistry are used to characterize the different combustion modes emerging in mixing layers separating nitrogen-diluted counterflowing planar streams of hydrogen and oxygen. Attention is focused on high degrees of dilution, resulting in near-limit flames, with peak temperatures close to the crossover temperature. A bifurcation diagram is presented in a plane, having the stoichiometric mixture fraction and normalized strain rate as coordinates, that identifies six different combustion regimes involving four different flame types, namely, diffusion-flame sheets, advancing and retreating edge flames, multiple flame tubes, and single isolated flame tubes. Multiple-tube flame configurations vary from small, round, widely separated flame strings at high strain rates to wide, flat, densely packed flame strips, with narrow flame-free gaps between them, at lower strain rates, and they are steady and stable in various arrays over a continuum of tube-separation distances. The observed flame behavior exhibits hysteresis in a certain range of parameters, with the structure that is established depending on the ignition mechanism, as it also does at high strain rates, and a continuum of different stable steady-state flame configurations exists, each accessed from a different initial condition.
\end{abstract}


\end{frontmatter}

\section{Introduction}
\label{introduction}

Criticality conditions in hydrogen-oxygen combustion are fundamentally related to the existence of a chemical-kinetic crossover temperature $T_c$ at which the effective rates of high-temperature chain branching, associated with the shuffle reactions, and chain recombination, mainly controlled by the three-body reaction H+O$_2$+M~$\rightarrow$ HO$_2$+M, are equal~\cite{sanchez2014recent}. Chain branching dominates when the temperature in the reaction region lies above crossover, leading to the existence of a well-populated radical pool that maintains a high fuel oxidation rate, with characteristic times on the order of 1 ms. This branching path is no longer active below crossover, where chain-terminating processes are dominant. In this low-temperature regime, sustained hydrogen combustion relies instead on an extremely slow branching path involving HO$_2$ and H$_2$O$_2$, resulting in negligibly small oxidation rates. This is known to have a dramatic effect on ignition processes, for example, with associated induction times increasing to very large values (exceeding one second at $p=1$ atm) when the initial temperature falls below crossover. Because of this drastic change of reactivity, H$_2$-O$_2$ premixed and nonpremixed flames can exist only when the reaction-layer temperature lies above $T_c$ which thus determines the flammability limit.

Since branching and recombination tend to balance as the peak temperature approaches $T_c$, the effective rate of hydrogen oxidation becomes small in near-limit flames. The resulting weakly reactive solutions are affected by preferential-diffusion interactions that lead to known diffusive-thermal instabilities, which play a central role in the dynamics of near-critical H$_2$-O$_2$ combustion. For instance, they are responsible for the emergence of pulsating planar deflagrations in rich H$_2$-O$_2$ mixtures and for the formation of cellular flames in very lean H$_2$-O$_2$ mixtures \cite{lewis2012combustion}. In the latter case, the resulting curved flames, approaching flame-ball structures as limiting solutions \cite{fernandez2012flammability}, display superadiabatic flame temperatures, so that they can exist under conditions where planar deflagrations cannot, thereby extending significantly  the lean flammability limit. A wide variety of similar dynamic phenomena has been recently observed~\cite{zhou2019effect} in experiments of H$_2$-O$_2$-N$_2$ nonpremixed flames in a slot-jet facility when the conditions of dilution and strain produce peak temperatures approaching $T_c$. These conditions are to be further investigated in the present paper with the objective of providing a more complete characterization of the dynamical behavior of nonpremixed H$_2$-O$_2$ combustion through additional experiments and numerical computations.

The new results presented below, exploring extended ranges of strain rate, stoichiometry, and dilution, will be related to those of previous numerical analyses of differential-diffusion effects in counterflow nonpremixed systems \cite{L1,thatcher2000edges,short2001edge,L2,thatcher2002oscillatory}. These previous studies considered reactants with general Lewis numbers undergoing an irreversible reaction with a rate having unity reaction orders and a power-law \cite{thatcher2000edges,thatcher2002oscillatory} or Arrhenius \cite{L1,short2001edge,L2} temperature dependence. Instead, our computations will focus directly on the specific chemistry and transport properties pertaining to H$_2$-O$_2$-N$_2$ mixtures, as is needed to enable quantitative comparisons with the experimental measurements for given conditions of reactant-feed composition and strain to be made. Exploratory computations using a chemical-kinetic mechanism with 12 elementary reactions, which has been shown to provide excellent accuracy for most combustion conditions \cite{boivin2013four}, revealed that for the nonpremixed flames investigated here the effects of reactions involving H$_2$O$_2$ and of the initiation steps H$_2$+O$_2$~$\stackrel{6b}{\rightarrow}$ HO$_2$+H  and H$_2$+M~$\stackrel{9b}{\rightarrow}$ H+H+M are negligibly small, so that the needed chemistry description reduces to the nine elementary reactions shown in Table~\ref{skeletal}. As shown recently~\cite{fernandez2019one}, an even simpler chemistry description can be developed sufficiently close to crossover by exploiting the fact that all radicals obey a steady-state approximation, so that hydrogen oxidation can be described with a single overall reaction, whose rate can be expressed in explicit form in terms of the temperature and the reactant concentrations. The present investigation focuses on near-extinction highly strained flames, for which steady-state approximations tend to deteriorate, as demonstrated recently~\cite{weiss2019accuracies}. For this reason, although the one-step reduced mechanism is in principle well-suited for the near-limit flames investigated here, for increased reliability the computations presented below will employ instead the complete short mechanism of Table~\ref{skeletal}.

\begin{table}[h]
\centering
\begin{tabular}{l l l l l r }
\hline\hline
  & Reaction & & $B$ & \multicolumn{1}{l}{$n$} & $T_a$ \\ \hline
1f& H+O$_2$ $\rightarrow$ OH+O              & & 3.52  10$^{16}$  \quad &  -0.7       & 8590 \\[-0.15ex]
1b& OH+O $\rightarrow$ H+O$_2$              & & 1.05  10$^{14}$  \quad &  -0.313     & 132 \\[-0.15ex]
2f& H$_2$+O $\rightarrow$ OH+H              & & 5.06  10$^4$         & 2.67       & 3165 \\[-0.15ex]
2b& OH+H $\rightarrow$ H$_2$+O              &  & 2.94  10$^{4}$  & 2.64 & 2430  \\[-0.15ex]
3f& H$_2$+OH $\rightarrow$ H$_2$O+H   &  & 1.17 10$^9$         & 1.3        & 1825 \\[-0.15ex]
3b& H$_2$O+H $\rightarrow$ H$_2$+OH   &  & 1.42  10$^{10}$  & 1.18 & 9379 \\[-0.15ex]
4f& H+O$_2$+M $\rightarrow$ HO$_2$+M$^{\rm a}$     & $k_{0}$ \quad  &   5.75 10$^{19}$ & -1.4 & 0.0 \\[-0.15ex]
                                                      & & $k_{\infty}$ \quad  &  4.65  10$^{12}$  & 0.44 & 0.0 \\ [-0.15ex]
5f& HO$_2$+H $\rightarrow$ OH+OH                & &  7.08  10$^{13}$    &  0.0  &  148  \\  [-0.15ex]
6f& HO$_2$+H $\rightarrow$ H$_2$+O$_2$  & &  1.66  10$^{13}$    &   0.0 &  414  \\[-0.15ex]
7f& HO$_2$+OH $\rightarrow$ H$_2$O+O$_2$    & &  2.89 $\times$ 10$^{13}$    &   0.0 & $-250$  \\ [-0.15ex]
&&& 4.50 10$^{14}$    &   0.0 & $5500$  \\ [-0.15ex]
8f& H+OH+M $\rightarrow$ H$_2$O+M$^{\rm b}$ & & 4.00 $\times$ 10$^{22}$       & -2.0        & 0.0  \\[-0.15ex]
9f& H+H+M $\rightarrow$ H$_2$+M$^{\rm c}$   & & 1.30  10$^{18}$       & -1.0         & 0.0  \\[-0.15ex]
\hline \hline
\end{tabular}
\caption{Rate coefficients in Arrhenius form $k=BT^n\exp{(-T_a/T)}$ for the skeletal mechanism with rate parameters in mol, s, cm$^3$, kJ, and K. \newline
{\footnotesize $^{a}$Chaperon efficiencies: H$_2$ (2.5),  H$_2$O (16.0), CO (1.2), CO$_2$ (2.4), Ar and He (0.7), and 1.0 for all other species; Troe falloff with $F_c=0.5$}\newline
{\footnotesize $^{b}$Chaperon efficiencies: H$_2$ (2.5),  H$_2$O (12.0), CO (1.9), CO$_2$ (3.8), Ar and He (0.4), and 1.0 for all other species.}\newline
{\footnotesize $^{c}$Chaperon efficiencies: H$_2$ (2.5),  H$_2$O (12.0), CO (1.9), CO$_2$ (3.8), Ar and He (0.5), and 1.0 for all other species.}\newline
}
\label{skeletal}
\end{table}

\section{Formulation}
\label{formulation}

We consider the mixing layer separating two counterflowing slot jets at temperature $T_o=298$ K. The oxidizer and fuel streams are assumed to be a mixture of O$_2$ and N$_2$ with oxygen mass fraction $Y_{{\rm O_2}o}$ and a mixture of H$_2$ and N$_2$ with fuel mass fraction $Y_{{\rm H_2}o}$. In nonpremixed combustion the reactants reach the reaction layer in stoichiometric proportions, so there is interest in relating the composition of the feed streams to the composition of the stoichiometric gas mixture obtained by combining one unit mass of the fuel stream with $8 Y_{{\rm H_2}o}/Y_{{\rm O_2}o}$ units of mass of the oxidizer stream, resulting in a mixture with volumetric proportionality that can be written as H$_2$:O$_2$:N$_2$=2:1:$N$. Following~\cite{zhou2019effect}, we thus choose to characterize the stoichiometry and dilution of the system in terms of the stoichiometric mixture fraction $Z_s=1/(1+8 Y_{{\rm H_2}o}/Y_{{\rm O_2}o})$ and a parameter $N$, defined as the ratio of the number of moles of inert to the number of moles of oxygen in a mixture formed by combining the fuel and oxidizer streams in stoichiometric proportions. These two quantities determine the boundary values of the reactant mass fractions according to
\begin{equation}
Y_{{\rm H_2}o}=\frac{1}{Z_s (9+7N)} \quad {\rm and} \quad Y_{{\rm O_2}o}=\frac{8}{(1-Z_s)(9+7N)}. \label{boundary_reactant}
\end{equation}
By modifying the relative dilution of the feed streams while keeping the value of $N$ constant one may vary the stoichiometric mixture fraction in the range
\begin{equation}
\frac{1}{9+7N} \le Z_s \le \frac{1+7N}{9+7N},
\end{equation}
where the lower and upper limiting values correspond to pure fuel ($Y_{{\rm H_2}o}=1$) and pure oxygen ($Y_{{\rm O_2}o}=1$), respectively. When $N$ is large, as it is near critical conditions, this range extends from nearly zero to nearly unity.

In the thermo-diffusive approximation of constant density $\rho$ and constant transport properties, adopted here to simplify the numerical description, the velocity in the mixing layer is given by the familiar stagnation-point flow solution $(v_x,v_y,v_z)=(0,-A y', A z')$ in terms of the strain rate $A$, independent of time. As in [5-9] the composition and temperature of the resulting reactive flow will be assumed to be independent of $z'$, thereby reducing the time-dependent conservation equations for the reactive species $i$ and energy to
\begin{align}
\frac{\p Y_i}{\p t}-y \frac{\p Y_i}{\p y} -\frac{1}{L_i} \left(\frac{\p^2 Y_i}{\p x^2}+\frac{\p^2 Y_i}{\p y^2} \right)&=\frac{\dot{m}_i}{\rho A} \label{Yieq}\\
\frac{\p T}{\p t}-y \frac{\p T}{\p y} -\left(\frac{\p^2 T}{\p x^2}+\frac{\p^2 T}{\p y^2} \right) &=-\frac{\sum_i h^o_i \dot{m}_i}{\rho A c_p} \label{Teq}
\end{align}
involving the temperature $T$ and the species mass fractions $Y_i$ as dependent variables and the dimensionless time $t=A t'$ and the dimensionless coordinates $x=x'/\sqrt{D_T/A}$ and $y=y'/\sqrt{D_T/A}$ as independent variables, with $D_T$ denoting the constant thermal diffusivity. In writing~\eqref{Teq}, the specific heat at constant pressure $c_p$ is assumed to be constant, a reasonably good approximation for highly diluted H$_2$-O$_2$ combustion systems. The expression $c_p=32\times 10^{-3} (3+N)/(36+28N)$ J/(kg K), which approximates the specific heat of a stoichiometric mixture H$_2$:O$_2$:N$_2$=2:1:$N$ at $T=1000$ K, is used in the integrations.

A Fickian description with constant Lewis number $L_i$ is adopted in the transport description, with the values $L_{\rm H_2}=0.29$, $L_{\rm O_2}=1.10$, $L_{\rm H_2O}=0.89$, $L_{\rm H}=0.18$, $L_{\rm OH}=0.71$, $L_{\rm O}=0.69$, and $L_{\rm HO_2}=1.08$ used in the computations shown below, as corresponds to flames diluted with nitrogen, whose mass fraction is determined from those of the reactive species with use of $Y_{\rm N_2}=1-\sum_i Y_i$. As anticipated in earlier work~\cite{zhou2019effect}, differential-diffusion effects will be seen below to be more pronounced in systems with larger stoichiometric mixture fraction $Z_s$, when the hydrogen diffusivity dominates reactant transport. Because of the relatively limited temperature range at near-critical conditions with feed streams at $298$ K, which also favors improved accuracy of the thermo-diffusive approximations, in addition Soret diffusion is neglected for simplicity.

The source terms in~\eqref{Yieq} and~\eqref{Teq} involve the mass production rates per unit volume $\dot{m}_i$ and the enthalpies of formation $h^o_i$. The former is evaluated with use of the nine-step mechanism shown in Table~\ref{skeletal}. The rate constants shown, including falloff for the reaction H+O$_2$+M~$\stackrel{4f}{\rightarrow}$ HO$_2$+M and a bi-Arrhenius expression for the reaction HO$_2$+OH $\stackrel{7f}{\rightarrow}$ H$_2$O+O$_2$, are taken from the San Diego mechanism \cite{SDmech}. The species concentrations are calculated without invoking the assumption of constant density by using the expression
\begin{equation}
C_i=\frac{p}{R^o T} \frac{Y_i/M_i}{\sum_i Y_i/M_i},
\end{equation}
where $R^o=8.314$ J/(mol K) is the universal gas constant, $M_i$ is the molecular mass of species $i$, and $p=1$ atm is the pressure.

A rectangular computational domain with boundaries $y=\pm y_{max}$ and $x=\pm x_{max}$ will be considered in the integrations. Associated boundary conditions for the temperature and reactant mass fractions at the upper and lower boundaries are $Y_{{\rm H_2}}=Y_{{\rm O_2}}-Y_{{\rm O_2}o}=T-T_o=0$ at $y=+ y_{max}$ and $Y_{{\rm H_2}}-Y_{{\rm H_2}o}=Y_{{\rm O_2}}=T-T_o=0$ at $y=- y_{max}$, as correspond to the conditions in the feed streams, all other reactive species having a zero mass fraction there. On the other hand, Neumann conditions $\p T/\p x=\p Y_i/\p x=0$ are used on the lateral boundaries $x=\pm x_{max}$, allowing for the existence of both one-dimensional and two-dimensional flames.

\section{Preliminary considerations pertaining to near-limit H$_2$-O$_2$ diffusion flames}

\begin{figure}
\centering
\includegraphics[width = 0.5\textwidth]{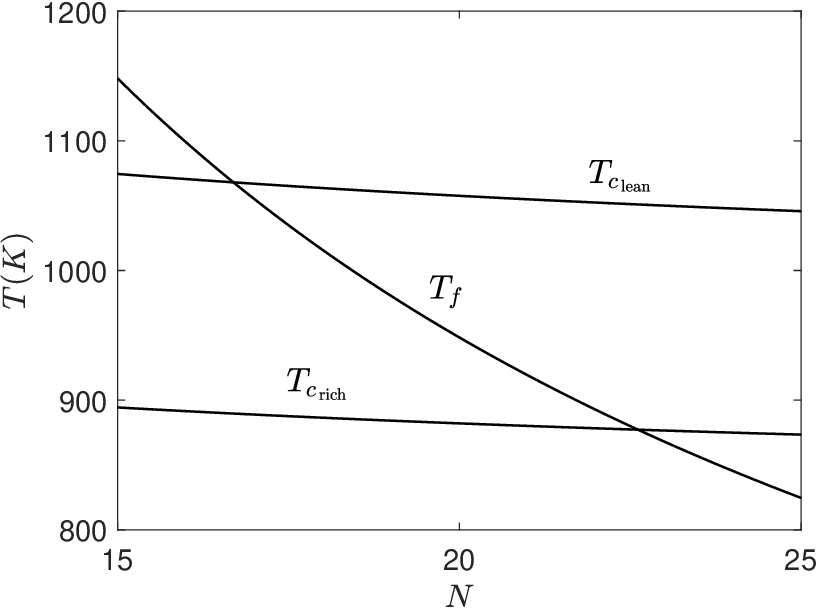}
\caption{The variation with the dilution parameter $N$ of the crossover temperature as obtained for $p=1$ atm from~\eqref{T_ceq} with $\alpha =1$ ($T_{c_{\scriptscriptstyle{\rm lean}}}$) and with $\alpha=k_{6f}/(k_{5f}+k_{6f})$ ($T_{c_{\scriptscriptstyle{\rm rich}}}$) with $C_{\scriptscriptstyle{\rm M_4}}=(32+N)/(2+N)$, and the premixed adiabatic flame temperature $T_f$ for an initial temperature of $T_o=298$ K.}
\label{fig:1}
\end{figure}

In hydrogen-oxygen combustion, the main chain-branching path at high temperature is established by the rate-controlling reaction H+O$_2$~$\stackrel{1f}{\rightarrow}$ OH+O followed by the faster O and OH consumption reactions O+H$_2$~$\stackrel{2f}{\rightarrow}$ OH+H and (twice)  OH+H$_2$~$\stackrel{3f}{\rightarrow}$ H$_2$O+H, resulting in an overall branching reaction 3H$_2$+O$_2$~$\rightarrow$ 2H$_2$O+2H with a rate $\omega_B=k_{1f} C_{\scriptscriptstyle{\rm O_2}} C_{\scriptscriptstyle{\rm H}}$. On the other hand, the main recombination path is initiated by H+O$_2$+M~$\stackrel{4f}{\rightarrow}$ HO$_2$+M, with HO$_2$ subsequently consumed through reactions HO$_2$+H~$\stackrel{5f}{\rightarrow}$ OH+OH, HO$_2$+H~$\stackrel{6f}{\rightarrow}$ H$_2$+O$_2$, and HO$_2$+OH~$\stackrel{7f}{\rightarrow}$ H$_2$O+O$_2$. Since $5f$ is chain-carrying, while $6f$ and $7f$ are chain-terminating, only a fraction
\begin{equation}
\alpha=\frac{k_{6f} C_{\scriptscriptstyle{\rm H}}+ k_{7f} C_{\scriptscriptstyle{\rm OH}}}{(k_{5f}+k_{6f}) C_{\scriptscriptstyle{\rm H}}+ k_{7f} C_{\scriptscriptstyle{\rm OH}}}
\end{equation}
of the HO$_2$ generated by H+O$_2$+M~$\stackrel{4f}{\rightarrow}$ HO$_2$+M ends up contributing to the overall recombination reaction 2H~$\rightarrow$ H$_2$, whose rate correspondingly becomes  $\omega_R=\alpha k_{4f} C_{\scriptscriptstyle{\rm M_4}} C_{\scriptscriptstyle{\rm O_2}} C_{\scriptscriptstyle{\rm H}}$. Equating the effective rates of H-atom production and consumption then leads to the equation
\begin{equation}
k_{1f}=\alpha k_{4f} C_{\scriptscriptstyle{\rm M_4}} \label{T_ceq}
\end{equation}
for computation of the crossover temperature $T_c$. The presence of $\alpha$ introduces a dependence of $T_c$ on composition, additional to that present through the effective third-body concentration $C_{\scriptscriptstyle{\rm M_4}}$, which incorporates increased chaperon efficiencies for H$_2$ (2.5) and H$_2$O (16.0), as seen in Table~\ref{skeletal}. In lean premixed flames and on the oxygen side of nonpremixed flames, where OH radicals are abundant, HO$_2$ predominantly reacts through HO$_2$+OH~$\stackrel{7f}{\rightarrow}$ H$_2$O+O$_2$, so that the recombination efficiency approaches $\alpha=1$. By way of contrast, in rich premixed flames and on the rich side of nonpremixed flames, where H radicals dominate the radical pool, HO$_2$ removal is controlled by  HO$_2$+H~$\stackrel{5f}{\rightarrow}$ OH+OH and HO$_2$+H~$\stackrel{6f}{\rightarrow}$ H$_2$+O$_2$, resulting in an overall recombination efficiency $\alpha=k_{6f}/(k_{5f}+k_{6f})$. For illustrative purposes, values of $T_c$ evaluated with $\alpha=1$ ($T_{c_{\scriptscriptstyle{\rm lean}}}$) and with $\alpha=k_{6f}/(k_{5f}+k_{6f})$ ($T_{c_{\scriptscriptstyle{\rm rich}}}$) are plotted in Fig.~\ref{fig:1}. As is appropriate in the reaction layer of the diluted hydrogen-oxygen diffusion flames considered here, the effective third-body concentration for step 4, $C_{\scriptscriptstyle{\rm M_4}}=(32+N)/(2+N)$, is evaluated in constructing the plot for a mixture of water vapor and nitrogen with mole fractions $2/(2+N)$ and $N/(2+N)$, corresponding to the equilibrium composition of a stoichiometric H$_2$:O$_2$:N$_2$ mixture with molar proportionality H$_2$:O$_2$:N$_2$=2:1:$N$.

As previously mentioned, near-critical flames appear under conditions such that the reaction-layer temperature lies close to the crossover temperature $T_c$, with the overall fuel-oxidation rate following closely a branching-recombination balance. For instance, in premixed combustion the condition that the adiabatic flame temperature must lie above the crossover temperature can be used to define the range of equivalence ratios in which a planar steady flame may exist for a given dilution. The computation of the corresponding lean and rich flammability limits must account for the existence of two distinct crossover temperatures in the two limits~\cite{fernandez2019one}. Similarly, in nonpremixed combustion a quick characterization of the reactant-feed conditions that lead to near-critical flames can be achieved by comparing $T_c$ to the peak temperature $T_p$ reached in the Burke-Schumann limit of infinitely fast reaction. In particular, a necessary condition for the existence of the diffusion flame is that $T_p>T_{c_{\scriptscriptstyle{\rm rich}}}$, whereas for $T_p$ much larger than $T_{c_{\scriptscriptstyle{\rm lean}}}$ near-critical phenomena can be expected to be of lesser importance.

\section{The planar diffusion-flame structure for an infinitely fast reaction}

In the Burke-Schumann limit, the flame appears as an infinitesimally thin flat surface, $y=y_f$ separating two equilibrium regions. Integrating the steady one-dimensional transport equations with boundary conditions $Y_{{\rm O_2}}-Y_{{\rm O_2}_o}=Y_{{\rm H_2}}=T-T_o=0$ as $y \rightarrow \infty$, $Y_{{\rm O_2}}=Y_{{\rm H_2}}-Y_{{\rm H_2}_o}=T-T_o=0$ as $y \rightarrow -\infty$, and $Y_{{\rm O_2}}=Y_{{\rm H_2}}=T-T_p=0$ at $y=y_f$ provides the reactant and temperature profiles
\begin{equation}
Y_{{\rm H_2}}=0, \quad \frac{Y_{{\rm O_2}}}{Y_{{\rm O_2}_o}}=1-\frac{1-\erf(y \sqrt{L_{{\rm O_2}}/2})}{1-\erf(y_f \sqrt{L_{{\rm O_2}}/2})}, \quad \frac{T-T_o}{T_p-T_o}=\frac{1-\erf(y/\sqrt{2})}{1-\erf(y_f /\sqrt{2})} \label{profiles1}
\end{equation}
for $y>y_f$ and
\begin{equation}
\frac{Y_{{\rm H_2}}}{Y_{{\rm H_2}_o}}=1-\frac{1+\erf(y \sqrt{L_{{\rm H_2}}/2})}{1+\erf(y_f \sqrt{L_{{\rm H_2}}/2})}, \quad Y_{{\rm O_2}}=0, \quad \frac{T-T_o}{T_p-T_o}=\frac{1+\erf(y/ \sqrt{2})}{1+\erf(y_f /\sqrt{2})} \label{profiles2}
\end{equation}
for $y<y_f$. The gradients on the oxidizer (superscript +) and fuel (superscript -) sides of the flame sheet are related by the jump conditions
\begin{equation}
\frac{(Y'_{{\rm O_2}})^+}{32 L_{{\rm O_2}}} =-\frac{(Y'_{{\rm H_2}})^-}{4L_{{\rm H_2}}} =\frac{(T')^+-(T')^-}{36 (h_{\rm H_2O}^o/c_p)} \label{jumps}
\end{equation}
obtained by integration of chemistry-free linear combinations of the steady one-dimensional conservation equations, with $h_{\rm H_2O}^o=-13.44 \times 10^6$ J/kg denoting the enthalpy of formation of water vapor.
The first equation in~\eqref{jumps}, stating that the reactants reach the flame sheet by diffusion with fluxes in stoichiometric proportions, can be evaluated with use made of~\eqref{profiles1} and~\eqref{profiles2} to give
\begin{equation}
\frac{1-\erf(y_f \sqrt{L_{{\rm O_2}}/2})}{1+\erf(y_f \sqrt{L_{{\rm H_2}}/2})}=\sqrt{\frac{L_{{\rm H_2}}}{L_{{\rm O_2}}}} \frac{Z_s}{1-Z_s} \exp[(L_{{\rm H_2}}-L_{{\rm O_2}}) y_f^2/2], \label{yf_eq}
\end{equation}
as an implicit equation for the flame location $y_f$. A similar evaluation of the second equation in~\eqref{jumps}, with use made of~\eqref{boundary_reactant}, yields
\begin{equation}
\frac{T_p-T_o}{T_f-T_o}=\frac{\exp[(1-L_{{\rm H_2}}) y_f^2/2]}{2 \sqrt{L_{{\rm H_2}}} Z_s}  \frac{1-\erf^2(y_f /\sqrt{2})}{1+\erf(y_f \sqrt{L_{{\rm H_2}}/2})} \label{Tp_eq}
\end{equation}
for the peak temperature $T_p$, where
 \begin{equation}
T_f=T_o+\frac{9}{9+7N} \frac{h_{\rm H_2O}^o}{c_p}, \label{Ta_exp}
\end{equation}
is the stoichiometric adiabatic temperature, whose variation with $N$ is plotted in Fig.~\ref{fig:1}.

\begin{figure}
\centering
\includegraphics[width = 0.5\textwidth]{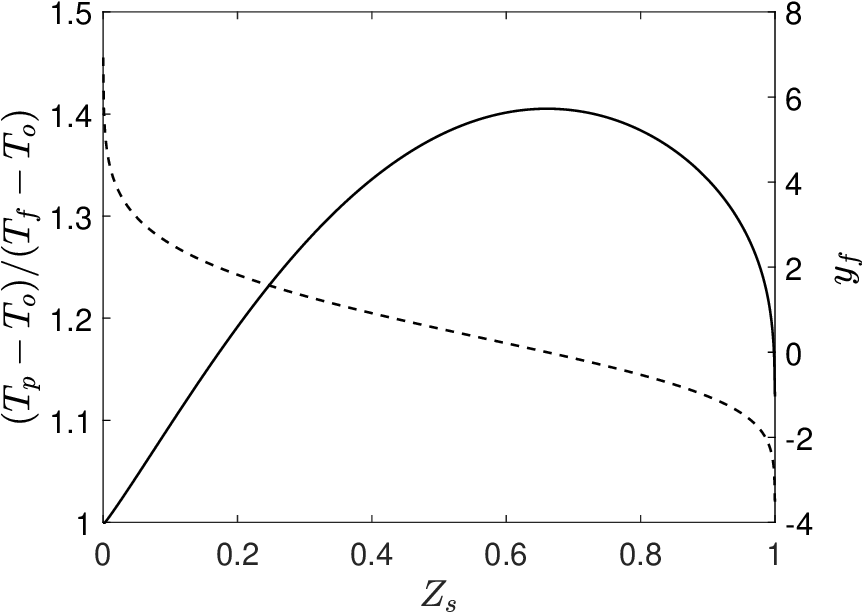}
\caption{The variation with $Z_s$ of the peak Burke-Schumann temperature (solid curve) and the reaction-sheet loctaion (dashed curve) of a counterflow diffusion flame as obtained from~\eqref{yf_eq} and~\eqref{Tp_eq} with $L_{\rm H_2}=0.29$ with $L_{\rm O_2}=1.10$.}
\label{fig:2}
\end{figure}

As can be seen from~\eqref{yf_eq}, in the equidiffusional case $L_{{\rm H_2}}=L_{{\rm O_2}}=1$ the flame location is given by $\erf(y_f/\sqrt{2})=1-2 Z_s$, which can be used in~\eqref{Tp_eq} to yield the familiar result $T_p=T_f$, independent of $Z_s$. Numerical evaluation of~\eqref{yf_eq} and~\eqref{Tp_eq} is in general needed to compute $y_f$ and $T_p$ for the realistic values $L_{{\rm H_2}}=0.29$ and $L_{{\rm O_2}}=1.10$, with results shown in Fig.~\ref{fig:2}. From this plot it can be seen that the temperature increase associated with differential-diffusion effects is more pronounced for intermediate values of $Z_s$, skewed towards larger values, enabling the existence of planar diffusion flames with levels of feed-stream dilution well above the value $N=23$ predicted from the simple adiabatic-flame criterion $T_f=T_{c_{\scriptscriptstyle{\rm rich}}}$. Additional differential-diffusion effects induced by curvature are to be investigated below in the following two-dimensional computations.

\section{Extinction strain rates of planar diffusion flames}

In general, in studies focusing on edge-flame propagation, one may, logically speaking, use the laminar planar premixed flame speed as the characteristic velocity scale for the edge-flame speed, provided that the effects of thermal expansion are ignored, since the classical tri-brachial structure approaches the planar flame as the strain rate becomes small compared with the extinction strain rate of the one-dimensional diffusion flame. Although this is also true for the current problem, the strain rate that is needed to achieve the triple flame is so small that it can never be realized in actual experiments; in the slot-jet apparatus, heat loss to the jet walls and other body-force effects, not considered in our formulation, will prevent the occurrence of these low-strain triple flames. Moreover, since interesting flame dynamics occurs near the extinction point for the near-critical limits, it is natural to use the one-dimensional extinction strain rate as the representative inverse time scale, as has been done previously~\cite{zhou2019effect}; that is, instead of normalizing predictions in terms of a characteristic velocity, it becomes better to normalize them in terms of a characteristic strain rate.

\begin{figure}
\centering
\includegraphics[width = 0.5\textwidth]{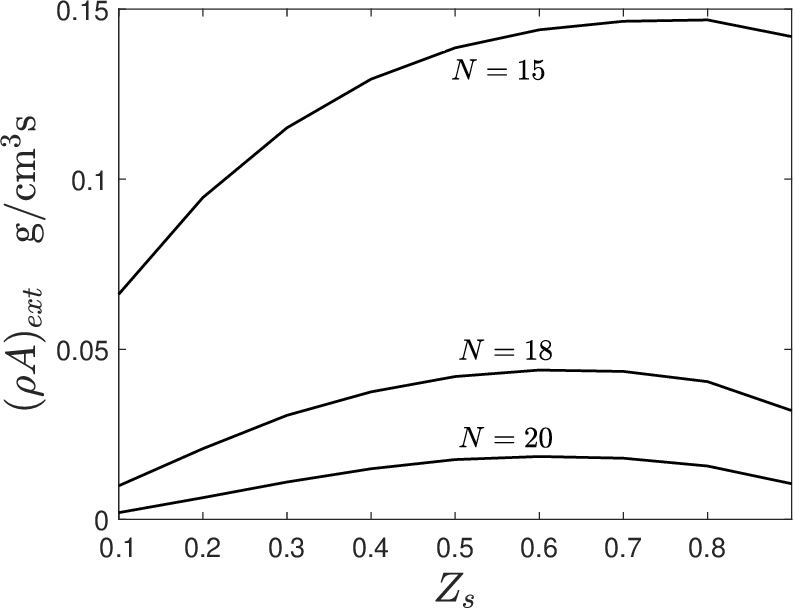}
\caption{The extinction value of the product $\rho A$ for the diffusion flame as a function of the mixture fraction $Z_s$ for three different values of $N$.}
\label{fig:ext}
\end{figure}

The extinction values of the strain rate are obtained from the maximum value of the product $\rho A$ for which equations~\eqref{Yieq} and~\eqref{Teq}, after neglecting $t$ and $x$ derivatives, have a solution. These values, denoted by $(\rho A)_{ext}$ and calculated as functions of $Z_s$ for three different values of $N$, are shown in Fig.~\ref{fig:ext}. Measuring the species-production rate $\dot{m}_i$, and the heat-generation rate $-\sum_i h^o_i \dot{m}_i$, in units of $(\rho A)_{ext}$ and $c_pT_o (\rho A)_{ext}$, the only parameter that appears in equations~\eqref{Yieq} and~\eqref{Teq} becomes $\ep = (\rho A)/(\rho A)_{ext}=A/A_{ext}$, the non-dimensional strain rate. In this manner, it is unnecessary to select any particular value for $\rho$ in presenting results of computations.

\section{Computational procedures}

Equations~\eqref{Yieq} and~\eqref{Teq} were integrated numerically, marching forward in time, subject to the stated boundary conditions, with different choices of the value of $x_{max}$ for application of the Neumann conditions, taken to be $x_{max}=25$ unless stated otherwise, and with $y_{max}=10$, approximating well a mixing layer of infinite extent because of the rapid decay with distance of disturbances moving against the increasing inflow velocities. The numerical method is based on a code developed for transient combustion applications~\cite{carpio2016local}. Integrations were performed for different selections of initial conditions, including a single hot spot or multiple hot spots (actually hot tubes because of the absence of a $z$ dependence\footnote{Flame tubes (which are not hollow) might instead be termed strips, stripes, strings, or rods, etc., but the selected term "tube" is most prevalent in the literature \cite{thatcher2000edges}.}), or a hot layer either extending to all $x$ or only from a chosen value of $x$ to the boundary of the domain, at a selected value of $y$. The evolution of the resulting structures with time was then computed with the strain-rate parameter $\ep$ held fixed, or else with that parameter varied slowly with time, to see whether hysteresis may occur as the system moves from one regime to another. In addition, to explore the range of possible steady solutions in regular arrays, the differential equations with the time-derivative terms removed were integrated in $-1 \le x/x_{max} \le 1$ with the independent variable $x$ replaced by $x/x_{max}$, so that $x_{max}$ appears as a parameter in the equations, and the range of values of $x_{max}$ for which such solutions exist was determined, the resulting flame configuration then becoming periodic in $x$, the length of the period being $2x_{max}$.

All of these computations were performed for the representative near-limit dilution value $N=18$, for which experimental results are available [4], since that is sufficient for exhibiting the full range of results, but the stoichiometric mixture fraction was varied over the complete range possible for this value of $N\, (0.007\, \text{to}\, 0.941)$ because the different regimes that are encountered depend strongly on the value of this mixture-fraction parameter.

\section{Flame configurations and structures}

\begin{figure}[ht]
\centering
\includegraphics[scale=0.6]{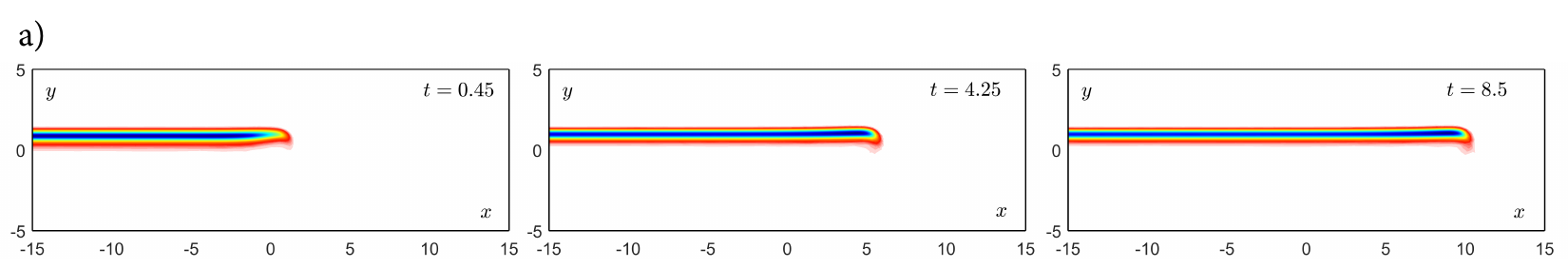} \\
\includegraphics[scale=0.6]{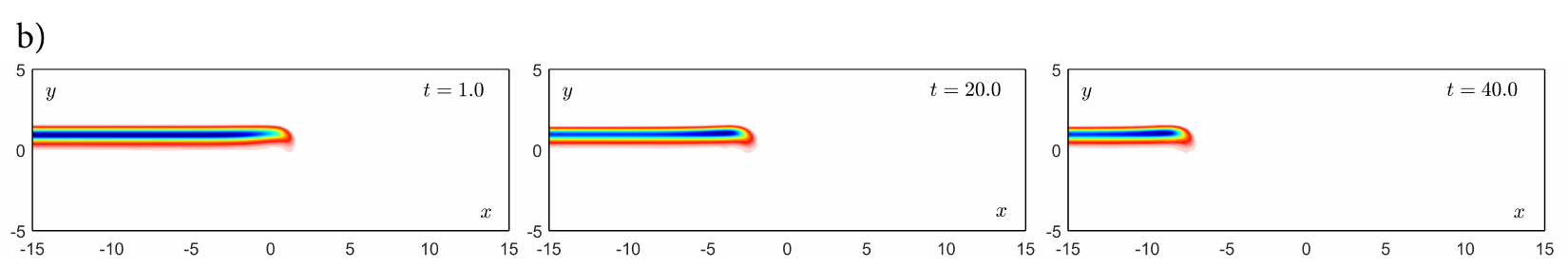}
\caption{Snapshots showing the evolution of the solution for $Z_s=0.4$ with $\ep=0.45$ (a) and $\ep=0.85$ (b)  when the integration is started with the flame extending only over a half-plane; flame shapes are visualized by color coding, red to blue with increasing computed OH mass fractions.}
\label{fig:Zs0.4}
\end{figure}

The flames that develop from different initial conditions differ for low and high values of $Z_s$. Representative computational results for low $Z_s$, when initially a flame extends over a half-plane, are those found at $Z_s=0.4$. Figure~\ref{fig:Zs0.4} shows the advancing (a) and retreating (b) edge flames encountered at lower strain rates ($\ep=0.45$) and higher strain rates ($\ep=0.85$), respectively, at that value of the stoichiometric mixture fraction. The OH mass fraction is selected as a representative measure of the flame structure. The region over which OH builds up is known to be broader on the oxygen side of the diffusion flame than on the hydrogen side for chemical-kinetic reasons, and that this is true can be inferred from the shading in the figure, especially at the flame edge, where downward bending of the initial buildup is discernible. At this value of $Z_s$, ignition always fails when $\ep>1$, extinction occurring irrespective of initial conditions.

\begin{figure}[ht]
\centering
\includegraphics[scale=0.6]{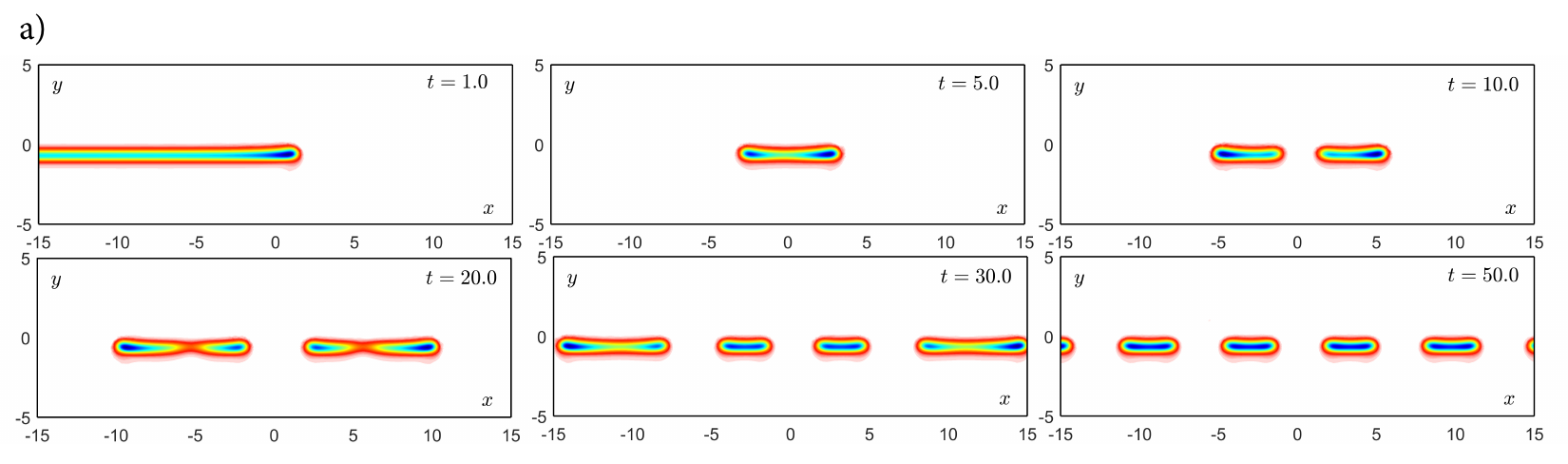} \\
\includegraphics[scale=0.6]{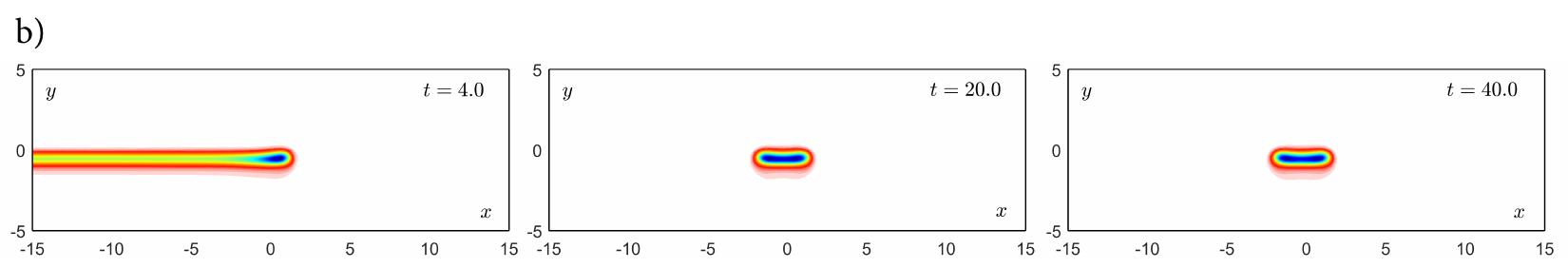}
\caption{Snapshots showing the evolution of the solution for $Z_s=0.9$ with $\ep=1.25$ (a) and $\ep=2.25$ (b)  when the integration is started with the flame extending only over a half-plane; flame shapes are visualized by color coding, red to blue with increasing computed OH mass fractions.}
\label{fig:Zs0.9}
\end{figure}

\begin{figure}[htbp]
\centering
\includegraphics[scale=0.6]{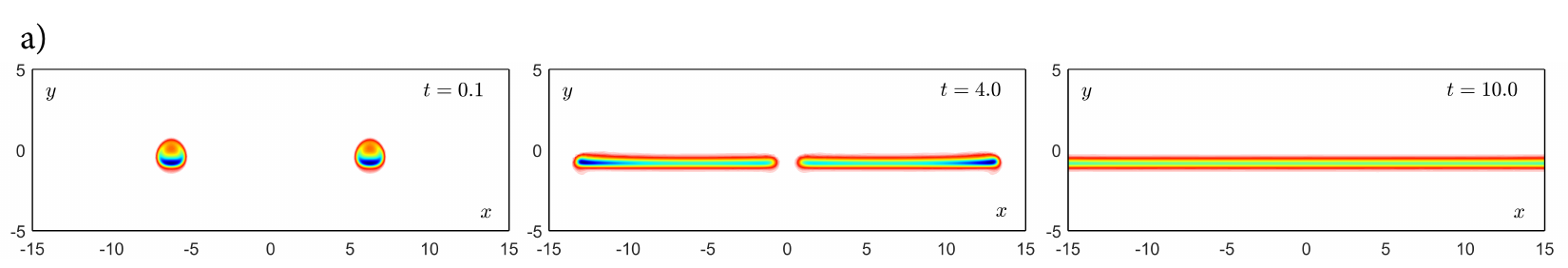} \\
\includegraphics[scale=0.6]{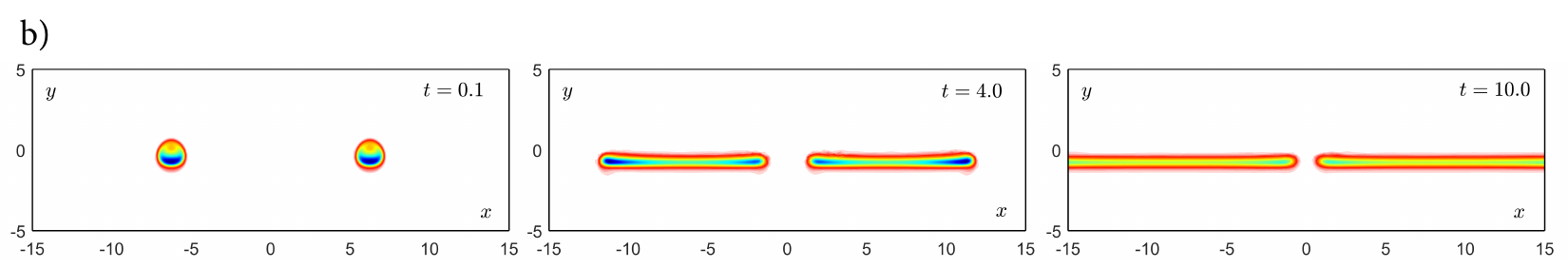} \\
\includegraphics[scale=0.6]{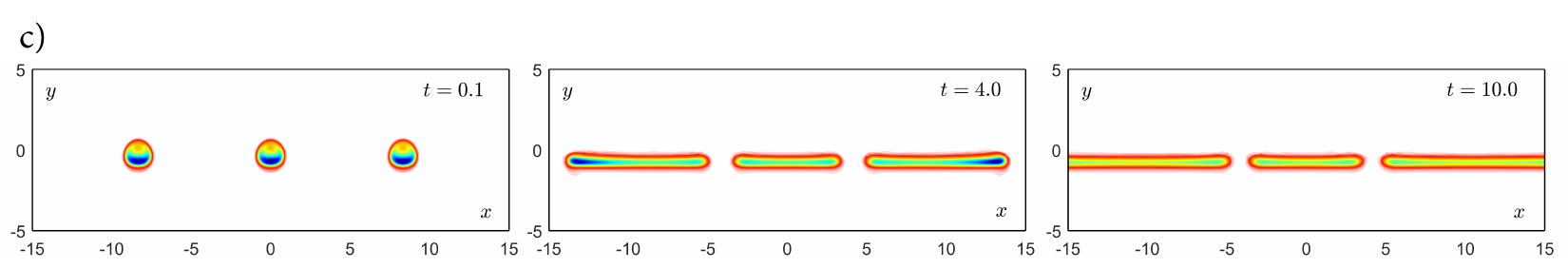} \\
\includegraphics[scale=0.6]{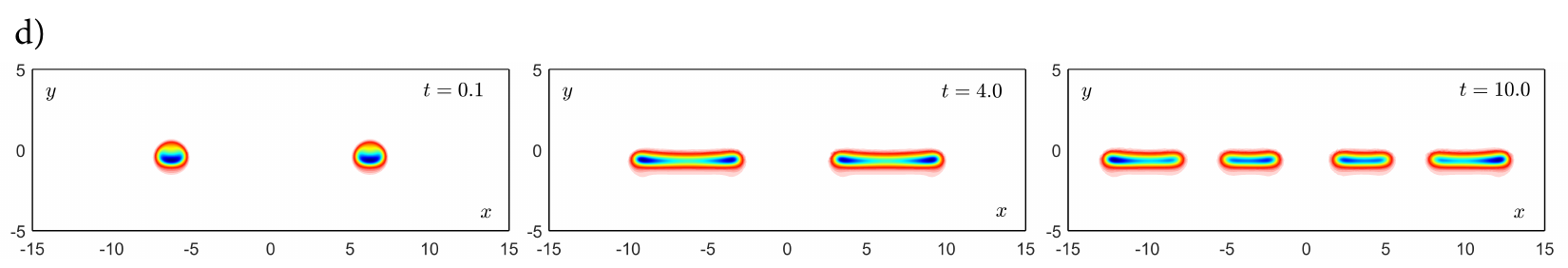} \\
\includegraphics[scale=0.6]{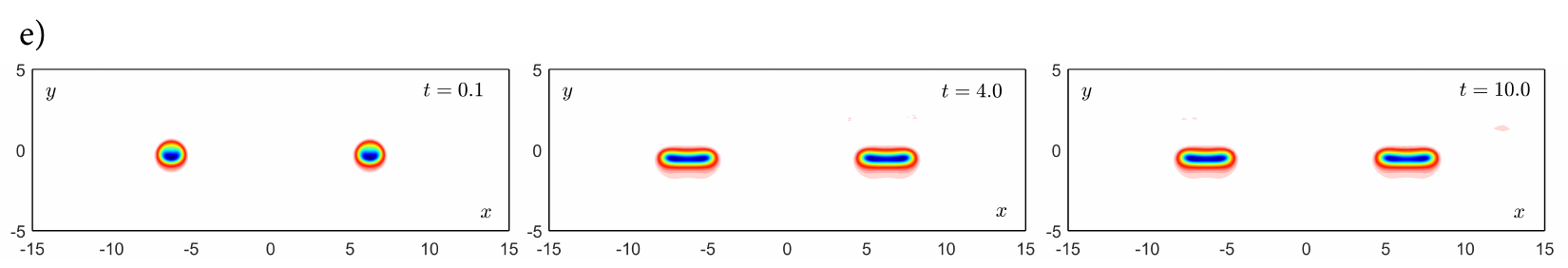}
\caption{Snapshots showing the evolution of the solution for $Z_s=0.9$ with $\ep=0.45$ (a), $\ep=0.70$ (b and c), $\ep=1.25$ (d), and $\ep=2.25$ (e)  when the integration is started with two or three hot spots; flame shapes are visualized by color coding, red to blue with increasing computed OH mass fractions.}
\label{fig:ignition}
\end{figure}

 At higher values of $Z_s$, however, the same half-plane initial condition results in continued combustion for some values of $\ep>1$, as can be seen for $Z_s=0.9$ in Fig.~\ref{fig:Zs0.9}. This figure shows that, at $\ep=1.25$, most of the half-plane extinguishes, but its end remains burning and develops into a flame tube, which subsequently splits in two, forming two flame tubes which, in turn, split again, finally leading to the array of tubes seen in the last panel of Fig.~\ref{fig:Zs0.9}(a). This splitting occurs because the relatively high chemical activity at the ends depletes the fuel concentration in the center, so that extinction occurs there. At the higher strain rate, $\ep=2.25$, however, the single flame tube formed by flame-sheet extinguishment fails to bifurcate and instead remains isolated, but nevertheless active, as seen in Fig.~\ref{fig:Zs0.9}(b). The higher strain rate reduces the chemical activity at the end so that the associated reduction in the nearby fuel concentration is insufficient to cause extinction at the center of the tube. At this value of $Z_s$, combustion can be initiated for strain rates all the way up to $\ep \simeq 3.5$, the sequence with increasing $\ep$ being similar to that in Fig.~\ref{fig:Zs0.9}(b), but with the final flame tube shrinking and becoming more round as $\ep$ increases.

Flame-tube dynamics and final flame-tube configurations can be sensitive to the initial conditions, especially at higher stoichiometric mixture fractions. Figure~\ref{fig:ignition} shows the evolution predicted for $Z_s=0.9$, with ignition produced by two or three hot spots, at four different values of $\ep$. At the lowest value shown, $\ep=0.45$, the complete planar diffusion flame is the only stable configuration, and the flame strips generated from the two hot spots are seen in Fig.~\ref{fig:ignition}(a) to merge and to form that planar flame. At the next higher value in the figure, $\ep=0.70$, the edges of the strips are unable to merge as a result of the opposing edges attracting hydrogen through preferential diffusion to a very large extent, thereby producing an insufficiently high hydrogen concentration in the region between them for combustion to occur there. Flame-free gaps thus persist between the burning regions, the widths of the gaps being determined by the strength of the preferential diffusion. That it is the gap width that is determined by the diffusion processes occurring may be inferred by comparing the sequences (b) and (c), the latter at the same condition but with three rather than two hot spots; the configuration evolves until the gaps that are present achieve the same width.  With the Neumann boundary conditions in the computations, the number of gaps remain one less than the number of initial hot spots.

This reactant-depletion effect of flame edges at low Lewis numbers, leading to the final configuration being controlled by the flame gaps, does not appear to have been recognized clearly in the previously cited publications. The implication, however, that regions exist in which there is more than one final steady configuration for given values of the strain rate and of the stoichiometric mixture fraction, that configuration being determined by the manner in which the steady solution is approached, is consistent with a previous finding \cite{L2} of multiple solutions. That work, which shares the essential low Lewis number of the present investigation, employed one-step Arrhenius chemistry, which demonstrates that the multiplicity phenomenon is not peculiar to hydrogen chemical kinetics.

The last two sequences in Fig.~\ref{fig:ignition} may be compared with Fig.~\ref{fig:Zs0.9} to see the differences between the evolutions that follow half-plane and multiple-hot-spot initial conditions. At $\ep=1.25$, instead of half-plane extinguishment, the hot spots grow into flame tubes, which subsequently continue to split, in the same manner as was observed for the tubes that resulted from half-plane ignition.  The final sequence shows that, while a single flame tube was generated by half-plane ignition, two result from ignition by two hot spots, yet another example of the multiplicity of steady solutions. The two-tube configuration is a requirement of the aforementioned inability of edge flames to merge above a critical value of $\ep$; it would be impossible for flame strips generated by two different hot spots to merge into a single flame tube. In the sequence (e) in the figure, the hot spots are far enough apart that they do not interact but instead independently develop into stable flame tubes; the number of flame tubes, in general, would be the same as the number of initial hot spots under these conditions.

These figures reveal various additional aspects of structures of the flame tubes. They appear to exhibit a continuous variety of shapes and structures. As extinction conditions are approached, they become almost entirely round strings, with the maximum reaction rate located at the center. Away from extinction, as conditions begin to admit more robust burning, they begin to flatten and develop reaction-rate peaks near their ends, those peaks having moved off from the central rate maximum. These aspects are reflected in the figures by the higher OH mass fractions near the ends. With increasing robustness, the tubes widen more, finally elongating into broad flame strips, before reaching a boundary at which they are transformed into edge flames that evolve to steady configurations controlled by the flame-free gaps between them. The peak temperatures in the flame tubes, which tend to exceed those in the strips, will be indicated below to decrease as conditions evolve towards boundaries at which flame extinction occurs. Flame-tube structures thus can be quite varied, depending on the strain rate and the stoichiometric mixture fraction. The broad range of flame-tube structures that are possible is remarkable.

\section{Evolution starting from a flame sheet or a flame tube}

\begin{figure}[ht]
\centering
\includegraphics[scale=0.65]{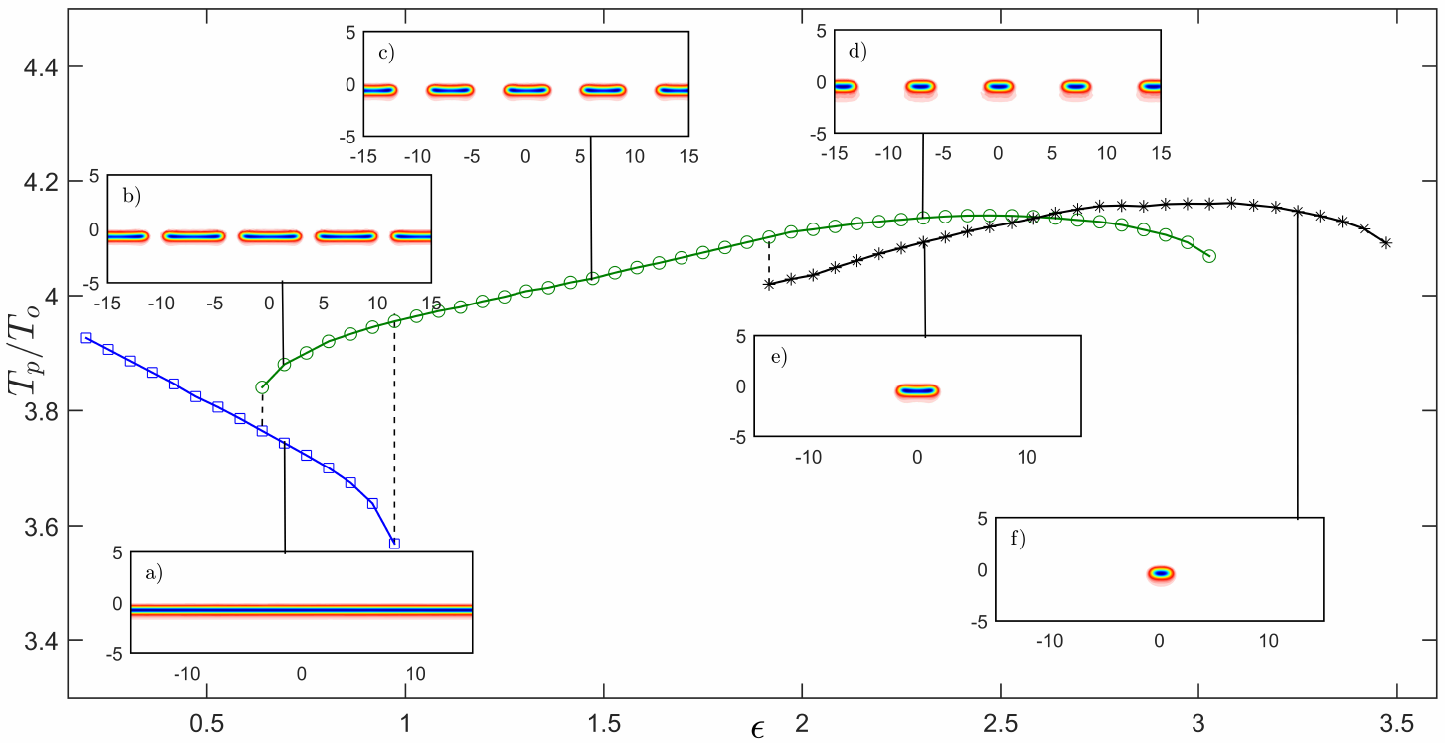}
\caption{Peak temperature as a function of $\ep$ for $Z_s=0.9$, obtained by slowly varying the strain rate, with resulting sample flame shapes visualized by color coding, red to blue with increasing computed OH mass fractions.}
\label{7}
\end{figure}

Figure~\ref{7} shows the computed variation of the peak flame temperature with the strain rate at $Z_s=0.9$, along with representative flame configurations and structures found in different regimes (shown by insets in the figure), that arise when continuous slow variations are imposed on the strain rate. The evolution develops as a consequence of the strong effects of preferential diffusion of hydrogen as the stoichiometric mixture fraction approaches unity at high dilution. The insets are identified by letters, for later reference.

Beginning at low strain rates at the far left, the decrease in the peak temperature with increasing strain rate up to $\ep=1$, seen in the square symbols, is associated with the infinite planar diffusion-flame structure illustrated by the color-coded image (a) in the lower left-hand part of the figure. This is consistent with the well-known decrease of flame temperature with increasing strain rate along the upper branch of the familiar S-shape curve of nonpremixed combustion~\cite{linan1974asymptotic}. When $\ep$ passes through unity, the planar flame might be expected to extinguish, resulting in only cold-flow mixing, but instead the planar flame was found to bifurcate, as is indicated in the figure by a vertical dashed line (rising up to the round symbols), experiencing a transformation into regularly spaced flame tubes, with structures similar to those which are illustrated in the last panel of sequence (a) in Fig.~\ref{fig:Zs0.9}.

Some time-dependent solutions just before $\ep$ reached unity showed the planar flame to be developing slight striations, as if it were going to break up into flame tubes, implying that the planar flame actually may become unstable to two-dimensional disturbances slightly prior to reaching planar extinction conditions. The corresponding difference in $\ep$, however, is so small that this effect, if it exists, is negligible. This finding, pertaining to hydrogen-oxygen chemistry, seems to be in agreement with previous diffusion-flame stability results corresponding to model chemistry~\cite{thatcher2002oscillatory,short2001edge}, for which the stability threshold is extremely close to the turning point of the S-shape curve. The principal conclusion to be extracted from these considerations is that the transition from the planar flame to the flame tubes for steady-state conditions is abrupt, resulting in a discontinuous increase of the peak temperature, which has been indicated by the vertical dashed line in the figure.

Clearly, therefore, for values of $\ep$ somewhat above unity, a stable steady combustion solution continues to exist, namely, a multiple-flame-tube solution, where the computation domain, with $x_{max}=25$, admitted seven tubes. The figure shows that, subsequent to the bifurcation, this multiple-tube steady solution experiences an increase of the peak temperature with increasing $\ep$, contrary to the dependence for the planar-flame solution. The opposite behavior is a consequence of the flame tubes becoming smaller and rounder with increasing $\ep$, which is indicated by the structures illustrated in the upper part of the figure (b, c, and d).

This increasing roundness is best explained by considering what happens to a round tube as the strain rate is decreased. A decrease in $\ep$ encourages edge-flame propagation, and so the round tubes tend to elongate. They elongate into flame strips that are more planar, thereby decreasing access to fuel in the flatter, central part of the flame tube; flame-tube elongation pulls the maximum temperature down towards that of the planar flame by decreasing the amount of heat release that can occur in the available residence time at the reduced ease with which fuel can reach the reaction zone. This tendency persists with increasing $\ep$ until the elongation effect disappears as the flame tube becomes sufficiently round. Subsequently, only a direct residence-time effect remains, and the resulting decrease in the residence time with increasing $\ep$ begins to cause a decrease in the amount of heat release, thereby decreasing the peak temperature, with increasing $\ep$. This residence-time effect persists in the round tubes until the peak temperature inside the flame tubes drops below the value needed to maintain chain branching over the reduced residence time available at the imposed strain rate, and extinction occurs.

 On the other hand, if, after the bifurcation at $\ep=1$, instead of being increased, the value of $\ep$ is decreased slowly, then stable, steady, multiple-tube solutions are found to continue to exist, almost until $\ep=0.6$, at which there is a jump to the planar solution, indicated by another vertical dashed line, with subsequent decreases of $\ep$ following the (now unique) planar-flame steady solution. The depletion of fuel in the gaps between the ends prevented merging of the flames from occurring until the strain rate was decreased to the value at the left-most vertical dashed line, where the depletion became insufficient. There thus is a region of hysteresis, where the planar flames is accessible only by increasing $\ep$ and the multiple flame tubes are accessible only by decreasing $\ep$, in the region between the two vertical lines. This type of hysteresis also was found previously with one-step Arrhenius chemistry \cite{L2}.

Instead of starting with a flame sheet, computations for slowly varying $\ep$ may be started with a single steady flame tube, such as that in the last panel of sequence (b) in Fig.~\ref{fig:Zs0.9}. The black curve, marked by asterisk points, that ends near $\ep=3.5$, pertains to a single flame tube in an infinite $x$ domain. This single-tube solution is the only one that exists in the limit as this maximum strain rate is approached, there being no steady flame-structure solutions of any kind for values of $\ep$ greater than that. The true extinction strain rate at $Z_s=0.9$ thus is about $3.5$ times that of the planar flame.

The single tube is almost perfectly round at the limit but elongates as $\ep$ is decreased (by the mechanism explained above), as is indicated by the two flame-structure insets (e) and (f) in the lower right-hand part of the figure. The two structures (d and e) shown for the same value of $\ep$, near 2.3, demonstrate the existence of two different steady flame-tube configurations under identical conditions, the solution with fewer tubes exhibiting greater elongation.

Starting at the extinction point, the single-tube peak temperature increases as the residence time available for reaction increases with decreasing $\ep$, but it reaches a maximum and thereafter decreases as the temperature reduction associated with tube broadening to approach a flame strip (the elongation effect) begins to become dominant. Eventually, near $\ep=1.9$, the single-tube solution ceases to exist; splitting into a multiple-tube solution, by the mechanism described in connection with part (a) of Fig.~\ref{fig:Zs0.9}, occurs at the short vertical dashed line. Splitting through gradual changes of $\ep$ is possible only between this and the next vertical dashed line at $\ep=1$, with the sheet splitting on the left and the single tube splitting on the right. While edge-flame propagation prevents splitting on the left, insufficient elongation prevents splitting on the right.

\section{Further deductions from steady-state computations}

\begin{figure}[ht!]
\centering
\includegraphics[scale=0.8]{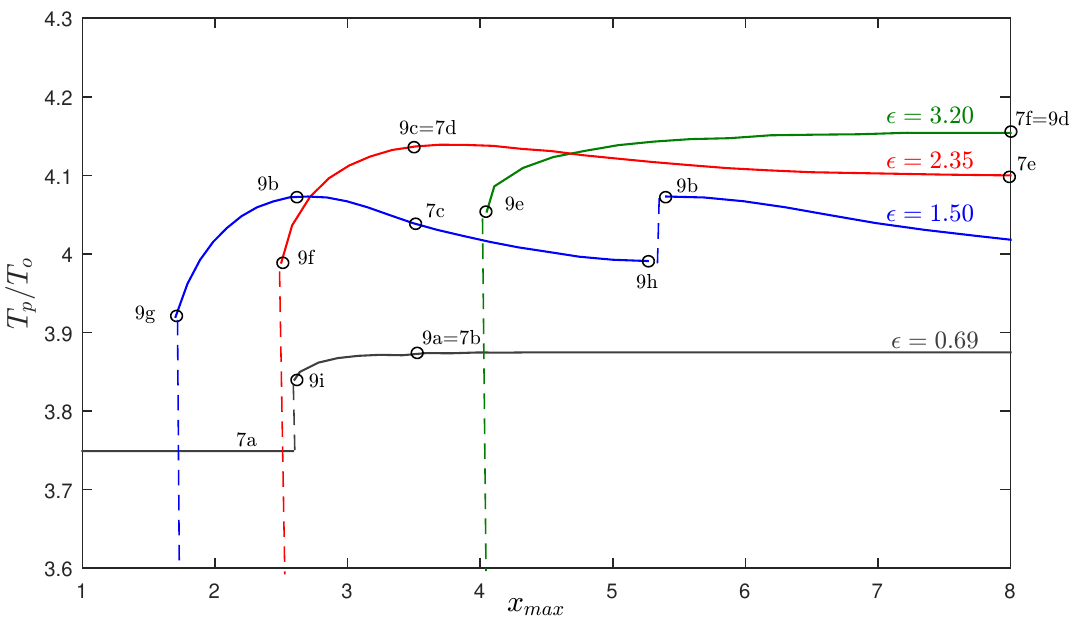}
\caption{Peak temperature as a function of $x_{max}$ for $Z_s=0.9$, obtained from steady-state computations for different values of $\ep$.}
\label{8}
\end{figure}

\begin{figure}[ht!]
\centering
\includegraphics[scale=0.65]{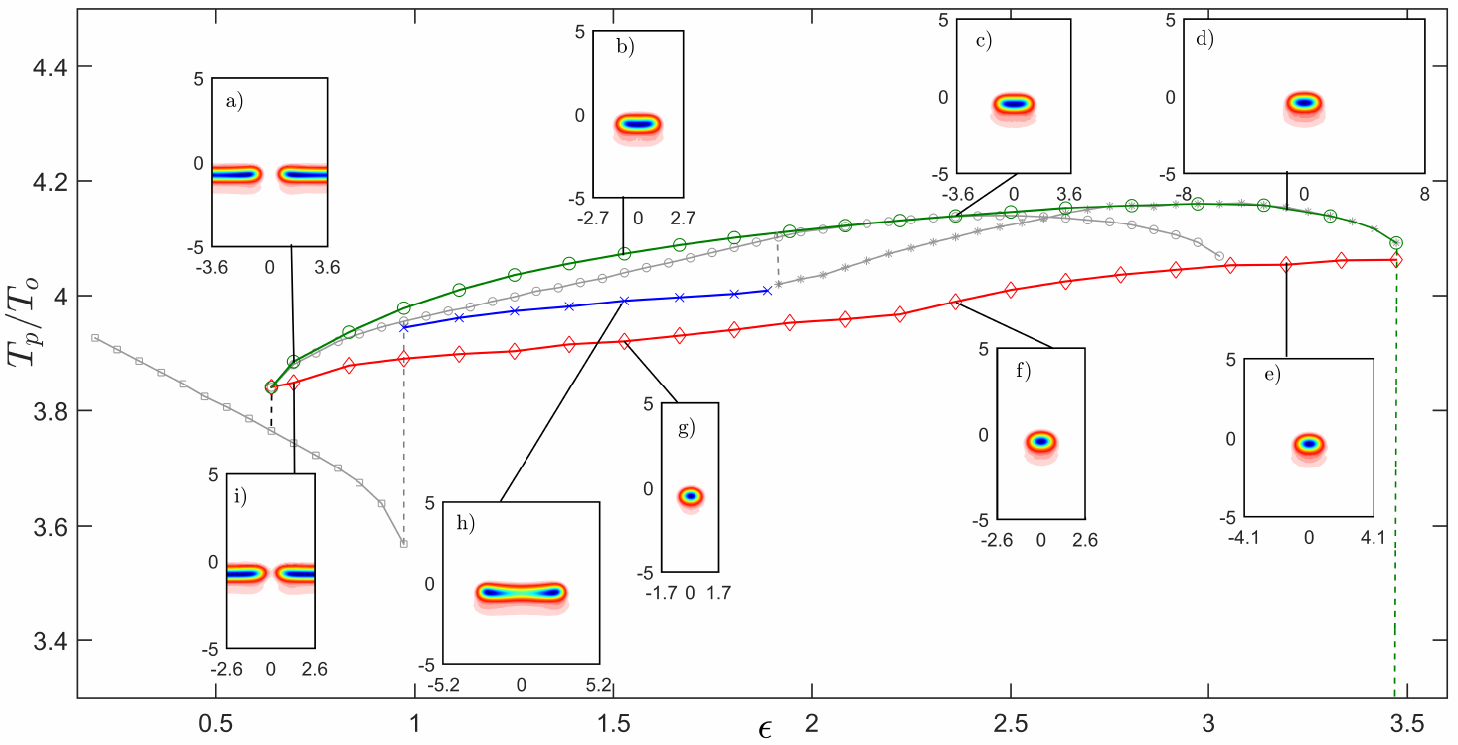}
\caption{Peak temperature as a function of $\ep$ for $Z_s=0.9$, obtained from steady-state computations treating $x_{max}$ as a parameter. Resulting sample flame shapes visualized by color coding, red to blue with increasing computed OH mass fractions.}
\label{9}
\end{figure}

The range of possible steady solution involving regular multiple-flame-tube arrays was explored in a separate set of computations. To that end, a new independent variable $-1\leq x/x_{max} \leq 1$ was introduced, with $x_{max}$ appearing as a parameter in the modified equations. Integrations, initiated with a flame tube located at the center of the computational domain, were performed by sequentially increasing and decreasing the value of $x_{max}$. Resulting peak temperatures (normalized by the ambient value) are shown in Figure~\ref{8} as a function of $x_{max}$ for four different values of the strain-rate parameter $\ep$, with the circles on the curves identifying conditions at which flame-tube structures are shown in Figs.~\ref{7} and~\ref{9}, the latter to be discussed below. For example, the three insets (b), (c), and (d) along the multiple-tube curve of Fig.~\ref{7}, corresponding to regular arrays of seven tube flames in a domain with $x_{max}=25$, are represented in Fig.~\ref{8} by the three circles $7$b, $7$c and $7$d corresponding to an inter-tube spacing $x_{max}=25/7=3.57$. Solutions associated with single flame tubes correspond to the limit in which the parameter $x_{max}$ approaches infinity, as occur with insets (e) and (f) of Fig.~\ref{7}, with corresponding temperature levels indicated by circles next to the right side of Fig.~\ref{8}.

For each one of the four strain conditions explored in Fig.~\ref{8} there exists a minimum value of the inter-tube spacing $x_{max}$, indicated by the circles $9$e, $9$f, $9$g, and $9$i, below which periodic arrays of flame tubes cannot exist. The resulting values are always larger than $x_{max}= 1$, a size of the order of the characteristic mixing-layer length. The values of $x_{max}$ at which the solution achieves the maximum temperature are also indicated along the curves by the circles $9$a, $9$b, $9$c, and $9$d. Since the single-tube solution is the solution in the limit of infinite wave length of the periodic multiple-tube solutions, its vanishing when $\ep$ is decreased below about $1.9$ means that, for smaller values of $\ep$, a maximum flame-tube separation distance must begin to exist, in addition to the minimum distance. This is seen in the curve for $\ep=1.5$, with the maximum value of $x_{max}$ marked with the circle $9$h. On the other hand, since elongation of the flame tube does not lead to flame splitting for $\ep<1$, the solution for $\ep=0.69$ does not exhibit a maximum value of $x_{max}$.

More extensive results of steady-state computations are summarized in Fig.~\ref{9}. For each value of the strain parameter $\ep$, the figure gives the maximum temperature of a multiple-flame-tube solution (large empty circles) along with the peak temperature of the tube array with minimum inter-tube spacing (diamonds) and with maximum inter-tube spacing (X symbols), the latter curve extending only in the intermediate range $1< \ep \ltsim 1.9$, as previously explained. To facilitate comparing results in different figures, the curves of Fig.~\ref{7} are reproduced by light gray color curves with tiny symbols. Figure~\ref{9} also shows insets of the solution at the values of $\ep$ shown in Fig.~\ref{8}. The insets are presented as the structure of just one of the multiple tubes in a uniform array, where the edge of each inset in this figure is a line of symmetry in the solution, so that the domain width of each inset is the separation distance in $x$ between adjacent flame tubes $2 x_{max}$.

The computations treating $x_{max}$ as a parameter show that, near the maximum value of $\ep$, namely $\ep=3.20$, steady stable solutions exist for uniform arrays of equally spaced multiple flame tubes with spacings ranging continuously from infinity to a minimum value. In other words, a continuously infinite number of steady-state flame-tube configurations exists under these near-extinction conditions! The inset (e) in the lower right-hand corner of Fig.~\ref{9} is shown to compare the flame-tube structure at the minimum separation spacing with that at maximum peak temperature for the same value of $\ep$, shown in inset (d) in the upper corner. The comparison of tube shapes in Fig.~\ref{9} at $\ep = 3.2$ shows that the neighbors have reduced the broadening of the tube by depriving it from receiving the fuel it would need to expand; its local atmosphere effectively is leaner, with the result that its peak temperature becomes lower as the flame-tube distance decreases, in agreement with the corresponding curve shown in Fig.~\ref{8}.

The light gray curve marked by tiny circles in Fig.~\ref{9} -corresponding to the curve marked by circles in Fig.~\ref{7}- represents multiple-tube arrays with inter-tube distance $x_{max}=25/7=3.57$. As expected, this curve ends on the right at $\ep \simeq 3.05$, when it reaches the minimum-spacing limit (diamond curve of Fig.~\ref{9}); for it to continue beyond that limit, the number of tubes in the fixed domain would have to decrease, which is impossible because flame-tube merging cannot occur when $\ep$ exceeds the left-most vertical line in Fig.~\ref{9} (a value of $\ep$ less than unity). Multiple-tube solutions with a continuous range of spacings exhibiting a minimum value continue to exist with decreasing $\ep$, all the way down to the farthest left vertical dashed line in Fig.~\ref{9}, where flame merging finally becomes possible, and the abrupt transition to the planar flame sheet occurs. The peak flame-tube temperature at this minimum separation decreases practically linearly with decreasing $\ep$, the lower strain rate allowing the round tube to survive to a lower peak temperature by reducing the rate of heat loss from the hot tube.

From Fig.~\ref{9}, with decreasing $\ep$, beginning around $\ep=2.7$, multiple-tube solutions exhibit peak temperatures above that of the single tube. This occurs when the shape-change effect of elongation begins to be dominant. The broadening of the single tube has reduced its peak temperature, so, by decreasing the broadening, adjacent tubes achieve higher peak temperatures than the single tube. But, as the spacing decreases further, the influence of adjacent tubes depriving the tube from its fuel begins to dominate, and the tube peak temperature then decreases, eventually below that of the single tube. This non-monotonicity is a complicating factor. The associated flame-tube shapes are shown by insets at $\ep=2.35$ in Figs.~\ref{7} and ~\ref{9}; the single tube in inset (e) of Fig.~\ref{7} is the broadest, and the tube with the minimum spacing, in inset (f) of Fig.~\ref{9}, is the smallest and roundest, but a tube with an intermediate spacing and shape (inset (c) of Fig.~\ref{9}, identical to one of the tubes in inset (d) of Fig.~\ref{7}) is the hottest.

The curve marked by X symbols, between the two central vertical dashed lines in Fig.~\ref{9} (i.e. for $1<\ep<1.9$), ending at the end of the single-tube curve, gives the peak temperatures of steady, stable tubes with maximum inter-tube spacing. The three flame-structure insets at $\ep=1.5$ in Fig.~\ref{9} show how the steady-state shape changes with tube spacing in that region. At the minimum spacing the tubes are practically round, as the spacing increases and reaches that needed to achieve the highest possible peak temperature there, the tube becomes more elongated, and, as the spacing reaches its maximum possible value, the tubes are extremely elongated and are about to split, as they can and do in this region. The entire possible range of steady flame-tube shapes thus is accessible in this region, the variations being achieved by varying the spacing in the regular array. Figure~\ref{8} for this value of $\ep$ show the dependence on the spacing from the smallest spacing to the spacing where splitting occurs; when the computations are continued to larger spacings the structure abruptly changes, the presence of two tubes reducing the spacing in $x_{max}$ from 5.2 to 2.6, essentially at the the maximum peak temperature, after which a further increase in the spacing reduces the peak temperature. Increasing the spacing beyond the range investigated would eventually lead to another splitting, with the pattern repeating indefinitely, but these are not new solutions, since they are merely reproducing solutions between the maximum-temperature spacing and splitting.

As $\ep$ is further decreased into the region of hysteresis, $\ep<1$, the maximum limit on the separation distance suddenly disappears, splitting no longer being possible, so the elongation becomes unbounded, with edges expanding until they encounter another edge. Since merging of edges, prevented by fuel depletion in the gap, is not possible until transition to the infinite planar flame sets in at the left-most vertical dashed line in Fig.~\ref{9}, the flame-free gap reaches a limiting width, and it is the gap width that controls the peak temperature (which occurs near the edge of the flame strip). The extent of fuel depletion in the gap will reach a maximum at the minimum separation distance because the depletion is enhanced by fuel diffusion to edges of adjacent gaps, so that, the closer gaps are to each other, the greater will be the depletion. The lower left-most inset (i) in Fig.~\ref{9}, at $\ep=0.69$, shown in a domain of width equal to the separation distance, is centered on the gap rather than on the flame because it is the gap that is controlling in this region. As the separation distance decreases, the flame tube ceases to exit, but it does not go to extinction; instead, it transitions to the infinite planar diffusion-flame solution, with the corresponding temperature level indicated by the horizontal line $7$a on Fig.~\ref{8}. As the separation distance increases, the associated decrease in the fuel-depletion influences of adjacent gaps allows the gap width to increase. The maximum extent of increase of gap size, which would correspond to the maximum value of the peak temperature, would occur in the limit of infinite separation distance (for $x_{max} \ltsim 3.5$ in the case $\ep=0.69$). The flame-structure inset in the upper left-most inset (a) in Fig.~\ref{9}, for the maximum peak flame temperature in this region, shows this maximum gap width; the flames at the sides of this inset would extend outward to infinity. Figure.~\ref{8} shows that rather small increases in the separation distance are sufficient to essentially increase the peak temperature to its maximum value. Actually, as the separation distance increases, it soon becomes large enough that further increases have negligible influences on the peak temperature, so it is practically independent of separation beyond the limiting separation that exhibits the maximum peak temperature for $\ep>1$, as is illustrated in Fig.~\ref{8}.

More dramatic is the discontinuous flame shape for the minimum separation distance as the strain rate decreases past $\ep=1$; because of the onset of the absence of any possibility for splitting, instead of the nearly round strings seen in the other insets, this limiting flame tube becomes a fairly wide strip, much like that which can be inferred from the inset (i) at the lower left in Fig.~\ref{9}; the shapes for insets (i) and (g) in this figure are quite different. Discontinuities of flame shapes also occur at this value of $\ep$ for separation distances greater than the minimum, further emphasizing that the importance of $\ep=1$ extends beyond the fact that the planar flame extinguishes there. It is the boundary conditions that force the variety of regular arrays in this range; without them the outermost edges would expand to infinity, forming semi-infinite flame sheets.

A further relevant observation is that, in this hysteresis region $\ep<1$, there exist infinitely many more steady, stable ``flame-tube" configurations than exist at higher strain rates. In particular, arrays need not be regular; any initial arrangement of ignition spots or flame-sheet gaps will experience edge propagation to an ultimate gap size dictated by locations of adjacent gaps. In irregular arrays, the maximum peak flame temperature will be controlled by the pair of adjacent gaps having the greatest separation distance between them. The range of values over which the peak temperature varies as the gap size is changed decreases as $\ep$ decreases, approaching a single value just before bifurcation, so that the jump in the peak temperature is the same, irrespective of which of the infinite number of possible stable configurations the bifurcation originates from. The bifurcation of the flame sheet at $\ep=1$, on the other hand, could occur only to a regular array of structures, but with any one of a continuously infinite number of spacings, between a maximum and minimum value, depending on the manner in which $\ep$ is increased; near-maximum spacings would be favored because their higher peak temperatures make them more robust.

\section{The regime diagram}
\label{bifurcation}

\begin{figure}
\centering
\includegraphics[width = 0.9\textwidth]{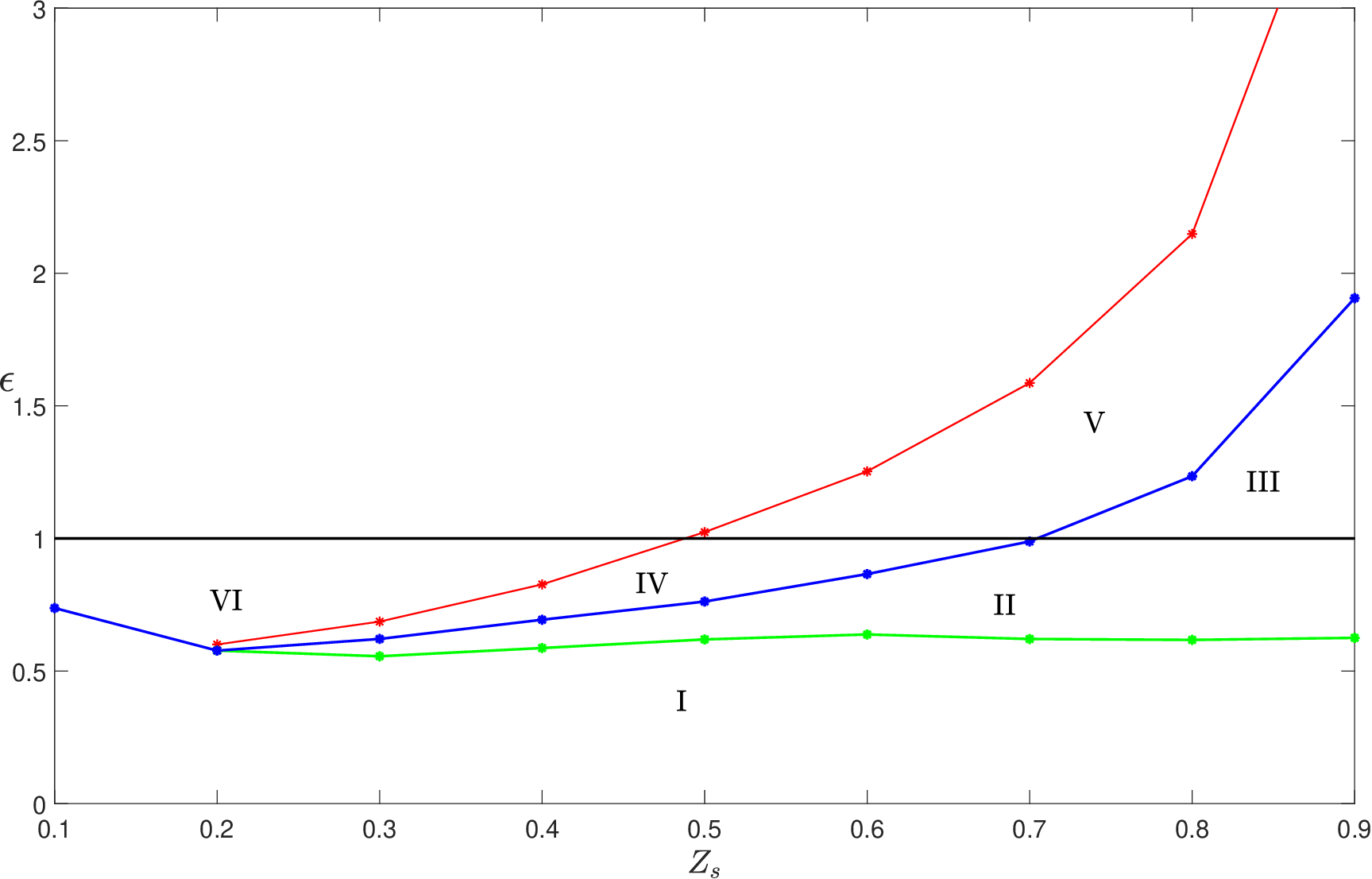}
\caption{The regime diagram, with stoichiometric mixture fraction and scaled strain rate as coordinates.}
\label{fig:regime}
\end{figure}

Computations were made for four different types of initial conditions, namely, combustion extending over a full plane, combustion extending only over a half plane, multiple isolated hot spots, as could be provided by multiple sparks, and a single isolated hot spot. Depending on the values of $Z_s$ and $\ep$, the combustion ultimately either quenched or evolved into one of four different types of configurations, namely, a full burning diffusion-flame plane, a burning half-plane, corresponding to either an advancing or retreating edge flame (in regions divided by a curve along which the edge flame remains stationary), multiple flame tubes that were either moving or stationary, or a single stationary flame tube. This variety of outcomes served to divide the $Z_s$ - $\ep$ plane into seven different regions, one without combustion at the end of the computation and six exhibiting continued combustion having different combinations of possible ultimate burning behaviors, with the boundaries between them separating the different flame configurations. This regime diagram is shown in Fig.~\ref{fig:regime}, where the six combustion regimes are labeled by Roman numerals, with boundaries color-coded for later reference.

The simplest boundary in the diagram is the horizontal (black) line $\ep = 1$, below which full planar steady-state diffusion-flame sheets exist, and above which they do not. The next most significant boundary is composed of three curves in the diagram, namely, the blue boundary between regions I and VI, followed by the red boundaries, first that between regions VI and IV, then the upper boundary of region V; flame tubes cannot exist above this boundary, where the only burning solutions are full stable planar diffusion flames or retreating edge flames, but they can exist below the part of it that begins at the left-most end of region IV where the red boundary begins. It is noteworthy that, for large enough values of $Z_s$, this boundary extends combustion into regions where all planar flames would extinguish, thereby widening the range of combustion through effects of preferential diffusion. A third notable boundary sequence is that between regions IV and II, followed by the boundary between regions V and III (blue); below that boundary edge flames are advancing (when they exist) and flame tubes are splitting and advancing, while above it edge flames are retreating and flame tubes are stationary, so that, for example, stationary edge flames exist only along this boundary (the lower part of it). The last significant boundary sequence is that between regions I and VI, followed by the green boundary between regions I and II; edges of flame sheets or tubes can merge below this boundary, but merging cannot occur above it, the strain rate and preferential diffusion being too large to permit two edges to combine, even though the planar flame that could develop after merging may be stable.

It is of interest to enumerate the flame configurations than can exist in each of the six different combustion regimes identified in the $Z_s-\ep$ plane, labeled by the Roman numerals in Fig.~\ref{fig:regime}. Region $\rm I$, which is approximately given by the condition $\ep<1/2$, is where a unique, stable, reacting equilibrium solution covering a complete plane eventually develops from any ignition event, irrespective of its type. A hot-spot ignition, or a semi-infinite hot-layer ignition (not extending to all values of $x$) leads to edge-flame propagation, leaving the diffusion flame behind as the steady-state solution. Multiple hot spots also result in edge propagation with the edges of adjacent hot spots merging together, forming a steady diffusion flame.

In region $\rm II$, both infinite, planar, diffusion-flame and broken-flame solution are allowable. In this region, a semi-infinite hot-layer ignition results in an advancing edge flame that eventually reaches the full, planar diffusion-flame solution, as does a single hot-spot ignition, but with more than one hot-spot ignition, perhaps surprisingly, this does not occur because the edges of the hot spots are unable to merge. With two hot spots, the result, instead, is two edge flames facing each other and extending to the boundary of the numerical domain, and with more than two hot spots the final configuration is an array of three or more complete elongated flame strips, the outermost two extending to a domain boundary. These developments occur basically because, in this regime, flame edges are unable to merge as a result of the opposing edges attracting hydrogen through preferential diffusion to a very large extent, thereby producing an insufficiently high hydrogen concentration in the region between them for combustion to occur there. Flame-free gaps thus persist between the flame tubes, the widths of the gaps being determined by the strength of the preferential diffusion. With multiple hot-spot ignitions, at separation distances large enough that adjacent hot spots do not interact, in this region flame tubes (or edge flames) are generated from each ignition spot and remain stable, without extinguishing. The line between region $\rm I$ and $\rm II$ is the smallest value of $\ep$ for which flame tubes can be realized as a final configuration.

By definition, in region $\rm III$, steady, infinite, planar diffusion flames do not exist, since this region is bounded from below by the condition $\ep=1$. In this region both single hot-spot and semi-infinite hot-layer ignitions lead to oscillatory propagation of leading flame edges, behind which flame sheets break and leave flame tubes behind; sheets must break, since edge flames also cannot exist here. The oscillation develops because the edge advancement speeds up after breaking, then slows as the next break prepares to occur. As $Z_s$ increases, the number of flame tubes for a given length increases in region $\rm III$. Flame tubes can elongate and split in this region as a consequence of the advancing of their edges in opposite directions, eventually leaving insufficient fuel adjacent to their mid-sections to maintain combustion there. Multiple-point ignitions in this region exhibit the flame-free-gap phenomena described for region $\rm II$, for the same reason. Flame-tube dynamics in this region thus appears to be more complex than in any other region.

The lower limit of region $\rm IV$ is the boundary at which an edge flame stops propagating and starts retreating as $\ep$ is increased. It also is the boundary curve at which the leading flame tube, generated by splitting of a flame sheet having an advancing flame edge (initiated by a central ignition spot) stops propagating, so that the flame tubes become stationary in region $\rm IV$. Isolated flame tubes, in addition to flame-tube arrays, also can be observed in this region, which thus permits a number of different steady flame-tube configurations, in addition to the planar diffusion flame.

Region $\rm V$ has the same characteristics as region $\rm IV$, except that in this region there are no retreating edge flames and no planar diffusion flames, there being only flame tubes, which ultimately remain stationary in this region. In addition, the continued existence of single isolated flame tubes is possible only in regions $\rm IV$ and $\rm V$. It is perhaps curious that, while flame tubes can exist only in regions $\rm II$, $\rm III$, $\rm IV$, and $\rm V$, those in region $\rm II$ are stable only because flame-free gaps are stable there, unlike those in the other three of these regions.

There are no flame tubes in region $\rm VI$; the only steady solution there is the infinite planar diffusion flame. The curve that separates region $\rm I$ and $\rm VI$ is associated  purely with the distinction between advancing and retreating edge flames. Thus, there exists no flame-tube dynamics for small $Z_s$. Flame-tube dynamical phenomena are confined to the range $\ep<1$ if $Z_s<1/2$; they are then present only in a narrow band at these low stoichiometric mixture fractions. On the contrary, for $Z_s>1/2$, flame-tube dynamics can be realized for both $\ep<1$ and $\ep >1$ with their region of existence growing larger as the value of $Z_s$ is increased.

The range of conditions investigated numerically in Fig.~\ref{fig:regime} corresponds to the experimental results of flame propagation reported in Fig.~5b of~\cite{zhou2019effect}. Partial correspondence of regimes and parametric boundaries between both diagrams can be found for intermediate values of $\ep$, for which the experimental flow remains laminar with negligible heat losses. Thus, the solutions found in regions I, III, V, and VI in Fig.~\ref{fig:regime} correspond, respectively, to modes I (advancing continuous flames), III (advancing broken flames), IV (stationary broken flames), and II (retreating continuous flames) in Fig.~5b of~\cite{zhou2019effect}. More work is needed to delineate a complete correspondence between the numerical and experimental diagrams. Specifically, experimental characterization of regions IV and II in Fig.~\ref{fig:regime}, not identified in the previous flame-propagation experiments, would require consideration of spark-ignition events for intermediate values of $Z_s$ and moderate strain rates in the range $0 < 1-\ep \ll 1$. On the other hand, modification of the energy equation~\eqref{Teq} by addition of a heat-loss term would be needed to model flame extinction by heat losses to the combustor wall, found in~\cite{zhou2019effect} for sufficiently small values of $\ep \ll 1$.

It is also of interest to relate the present results to those of previous edge-flame modeling efforts pertaining to reactants with equal diffusivity undergoing an irreversible reaction \cite{thatcher2000edges,thatcher2002oscillatory}. In these previous works, the reactants Lewis number $Le <1$ and the Damk\"ohler number $\delta$ (the ratio of the characteristic reaction rate to the strain rate) were varied to investigate the conditions for existence of different edge-flame phenomena, including advancing and retreating fronts, as well as diffusion-flame breakup into multiple tubes, or isolated flame tubes. The resulting map of solutions, plotted in the $Le$-$\delta$ plane in Fig.~4 of \cite{thatcher2000edges} (or in Fig.~6 of \cite{thatcher2002oscillatory}), is fundamentally equivalent to the $Z_s-\ep$ regime diagram shown in Fig.~\ref{fig:regime}, the only exception being region II (non-merging tubes)  in Fig.~\ref{fig:regime}, which was not identified in the previous work. To relate the parametric maps it is necessary to take into account the fact that increasing values of $\ep$ in Fig.~\ref{fig:regime} correspond to decreasing values of the Damk\"ohler number $\delta$ in Fig.~4 of \cite{thatcher2000edges} as well as the fact that  the effective Lewis number of H$_2$-O$_2$-N$_2$ mixtures decreases with increasing $Z_s$ (because the deficient reactant switches from oxygen to hydrogen as $Z_s$ increases from $Z_s \ll 1$ to $1-Z_s \ll 1$). With that in mind, it is easy to see that the lines $\delta=\delta_q$, $\delta=\delta_o$, and $\delta=\delta_e$ in \cite{thatcher2000edges} correspond to the black, red, and blue boundaries in Fig.~\ref{fig:regime}.

\section{Conclusions}

Near-limit H$_2$-O$_2$-N$_2$ combustion in counterflow mixing layers, studied numerically here, exhibits a noteworthy variety of different types of flame dynamics, which can be categorized in a bifurcation diagram. In a number of regimes in the diagram, the resulting flame structures and configurations are not unique; they depend very much on the initial conditions and therefore on the way in which any particular corresponding experiment would be carried out. Even though the thermo-diffusive approximation is invoked in the analysis, the flame shapes derived are in qualitative agreement with those seen in experiments~\cite{zhou2019effect}, although further testing against future experiments certainly would be worthwhile.

In a classification based on the extent and connectivity of reaction regions, planar two-dimensional counterflows of hydrogen highly diluted by nitrogen, against oxygen highly diluted by nitrogen, possess an infinite number of different steady-state, stable, concentration-field configurations. One class involves simple non-reactive mixing. Another has a reaction plane of infinite extent, located at the stoichiometric position, oriented perpendicular to the direction of injection. A third is an edge flame occupying a half plane perpendicular to the direction of injection, at that same stoichiometric position.  A fourth is a single flame tube, with its axis in the plane of the flow, oriented parallel to the outflow direction, located at a stoichiometric position.

The variety of possibilities in  H$_2$-O$_2$-N$_2$ systems increases with increasing values of the stoichiometric mixture fraction. There are configurations with equally spaced flame arrays for sufficiently high stoichiometric mixture fractions, having a continuum of possible spacings in the array. This continuum of spacings exhibits a minimum possible spacing over a wide range of strain rates, from a minimum strain rate, below which the only stable steady flame is the planar intact flame sheet, to the maximum possible strain rate above which combustion cannot occur. There also is a maximum possible spacing, over a smaller strain-rate range, from the strain rate at which the infinite planar flame sheet quenches to the strain rate below which a single, isolated flame tube cannot exist. In addition, for sufficiently high stoichiometric mixture fractions there is a range at smaller strain rates exhibiting hysteresis, over which two types of stable steady-state solutions exist, one involving a single infinite planar flame and the other multiple flame strips in arrays that need not be regular, with a range of possible inter-strip spacings, these gap sizes controlling the peak flame temperature, the strip widths being irrelevant.

The flame tubes exhibit a continuum of shapes and structures. Above the strain rate at which the infinite planar flame extinguishes, as extinction conditions are approached they become practically entirely round strings, with the maximum reaction rate located at the center. Away from extinction, as conditions begin to admit more robust burning, they flatten and develop reaction-rate peaks near their ends, elongating into fairly broad flame strips with flame-free gaps between them, before reaching a boundary at which they are transformed into edge flames, where it becomes more relevant to focus on the sizes of the gaps between the edges than on the flame sizes. The peak temperatures in the flame tubes, which exceed those in planar flames, increase as flame strips become shorter but then decrease as conditions evolve towards boundaries at which flame extinction occurs. Flame-tube structures thus can be remarkably varied, depending on the strain rate and the stoichiometric mixture fraction. The extremely broad range of steady, stable flame structures that are possible likely would not have been anticipated in advance of this study.

It is fortunate that in practice, with the exception of certain designs, such as those involving fuel injection into recirculating exhaust gases at high EGR, stoichiometric mixture fractions remain sufficiently small that many of the flame-structure complexities of near-limit H$_2$-O$_2$-N$_2$ systems will not be encountered. Flame sheets, along with advancing and retreating edge flames, and perhaps some stationary flame tubes, may be anticipated under these low-stoichiometric-mixture-fraction conditions. An associated aspect, which could be detrimental in power-production applications but is beneficial from the viewpoint of flammability safety, is that the substantial increase of the extinction strain rate, above that of the planar flame, which requires a stoichiometric mixture fraction greater than 0.5, is absent.

Future computational, experimental, and theoretical research in this area would be of interest. It could be helpful to explore the extent to which an available systematic further reduction of the present chemistry to a one-step description can be successful. Experiments designed to achieve more than one flame configuration [11] under identical conditions with different ignition procedures also could be revealing. For example, a planar flame could be established under conditions where hysteresis is predicted, and holes could temporarily be blown into it, to see if flame tubes develop. Also, single and multiple sparks could be employed in ignition to investigate whether different numbers of stable flame tubes can be established. The wide variety of different phenomena encountered in this study points to a fertile field for future investigations.

\section{Acknowledgment}

We are especially indebted to Paul Ronney, whose extensive experimental work, including determination of a regime diagram, motivated this investigation. The work of JC was supported by the Ministerio de Ciencia, Innovaci\'on y Universidades through project No. PGC2018-097565-B-I00 and through travel grant No. CAS18/00426.

\bibliographystyle{elsarticle-num}

\bibliography{references}

\end{document}